# On the energy spectrum evolution of electrons undergoing radiation cooling

S. V. Bulanov[1,2], G. M. Grittani[1], R. Shaisultanov[1], T. Zh. Esirkepov[2], C. P. Ridgers[3], S. S. Bulanov[4], B. K. Russell[5], and A. G. R. Thomas[5]

[1] *Extreme Light InfrastructureI ERIC, ELI–Beamlines Facility, Za Radnici 835, Dolni Brezany 25241, Czech Republic*

[2] *National institutes for Quantum and Radiological Science and Technology (QST), Kansai Photon Science Institute, 8-1-7 Umemidai, Kizugawa, Kyoto 619-0215, Japan*

[3] *York Plasma Institute, Department of Physics, University of York, Heslington, York YO10 5DD, UK*

[4] *Lawrence Berkeley National Laboratory, Berkeley, California 94720, USA*

[5] *Gérard Mourou Center for Ultrafast Optical Science, University of Michigan, 2200 Bonisteel Boulevard, Ann Arbor, Michigan 48109, USA*

**Abstract**

Radiative cooling of electron beams interacting with counter-propagating electromagnetic waves is analyzed, taking into account the quantum modification of the radiation friction force. Central attention is paid to the evolution of the energy spectrum of electrons accelerated by the laser wake field acceleration mechanism. As an electron beam loses energy to radiation, the mean energy decreases and the form of the energy distribution also changes due to quantum-mechanical spectral broadening.

# 1. Introduction

It is well known that radiation friction can impose constraints on the highest attainable energy of charged particles accelerated by standard accelerators [1]. Additionally, it leads to radiative cooling of the accelerated electron beams and affects the electron beam emittance [2, 3]. Radiation losses determine the energy of cosmic rays accelerated in various objects in space [4-7], in particular, the energy of ultra-high energy cosmic rays. It also plays an important role in charged particle interactions with crystals [8]. The effects of radiative cooling on the dynamics of electrons interacting with strong electromagnetic waves have attracted significant attention, specifically in the interaction of high power laser radiation with matter [9-12]. These effects can be neglected in the case of relatively low radiation-intensity and small electron-energy. However, in the limit of extremely high wave-intensity, radiation friction effects dominate the dynamics of the charged particles [13-19] resulting in the radiation friction force approaching the strength of the driving force. As a result, the electron dynamics become highly dissipative with fast conversion of the electromagnetic wave energy to hard electromagnetic radiation. In the pulsar magnetosphere theory this regime called as "Aristotelian Electrodynamics" because due to radiation over-damping, the velocity–rather than the acceleration–of a charge is determined by the local electromagnetic field [20, 21]. For more information on Aristotelian physics, e. g. see [22] and literature cited therein. For laser radiation with a 1 μm wavelength, the radiation friction force modifies the electromagnetic wave interaction with matter at intensities above $I_R = 10^{23}\,\text{W/cm}^2$. Reaching this laser intensity will bring us to regimes that are almost completely unexplored experimentally. This will enable high efficiency generation of gamma flares, which is considered as one of the primary goals for high-power laser facilities [23-34].

Radiation friction plays a significant role in the acceleration of charged particles using lasers [33-36]. However, the above-mentioned intensity, $10^{23}\,\text{W/cm}^2$, corresponds to the interaction geometry where the electromagnetic wave simultaneously accelerates charged particles and provides the strong field for radiation reaction effects. Typical laser-target configurations for studies of these conditions include laser pulse irradiation of a solid foil and penetrating the overdense foil target [10]. The intensity of $10^{23}\,\text{W/cm}^2$ was demonstrated recently [37]. Another experimental configuration that allow for the study of radiation friction force effects is the collision of a high energy electron beam and a high intensity laser pulse. It was recently studied at the Gemini laser facility, with the results reported in [38] and [39]. Already at a moderate intensity of approximately $10^{20}\,\text{W/cm}^2$ significant radiation friction effects were observed.

It is well known that an electron beam colliding with a strong electromagnetic pulse and interacting with a strong magnetic field undergoes fast cooling and fast energy depletion [40-44], which changes the beam energy distribution. In regard to the study of strong field quantum electrodynamics [10, 45-47], it raises the question of whether an electron radiating its energy away during the interaction with a laser can reach the region of highest intensity with sufficient energy to make a number of different phenomena observable [48-52].

Below, we present the calculation and analysis of the electron energy distribution that results from an electron beam interacting with a strong counter-propagating electromagnetic wave. In the analysis, we take into account different initial (before interaction) energy distributions and quantum effects modifying the radiation friction force. We additionally address the evolution of the energy spectrum of electrons accelerated by the laser wakefield acceleration (LWFA) mechanism.

The article presents three aspects: being a brief overview of the previously published theory of the evolution of an electron beam during its interaction with a strong laser field, it also contains original results related to the stochastic behavior of a charged particle in the quantum limit, and also presents a collection of useful formulas and relationships for planning, carrying out and analysis of experiments on the interaction of ultrarelativistic electrons with laser radiation.

The paper is organized as follows. In Section 2 we retrieve the formalism used for description of the radiation friction force in the Landau-Lifshitz form with the quantum effects implemented in the form of the Gaunt factor. The equations from this Section are applied in the next Sections 3 and 4 for calculating the energy spectrum of the electron beam cooled in the interaction with strong electromagnetic pulse neglecting the stochastic broadening of electron distribution function. The electron spectrum evolution in this case presents momentum averaged energy distribution. In Section 5 electron energy spectrum broadening due to quantum mechanical stochasticity effects is described within the framework of the Fokker-Plank equation. In order to benchmark the Fokker-Planck equation approach used in Section 5 to account for quantum mechanical stochasticity effects, in Section 6, we present the results of a Monte Carlo simulation modeling electron beam collision with a laser pulse. Section 7 summarizes the conclusions.

## 2. Radiation Friction Force

In order to self-consistently describe the trajectory of an emitting electron, the Minkowski equations should be modified by adding the radiation friction force, $g^\mu$:

$$\frac{dp^\mu}{ds} = -\frac{e}{c} F^\mu{}_\nu u^\nu + g^\mu, \tag{1}$$

$$\frac{dx^\mu}{ds} = u^\mu. \tag{2}$$

The radiation friction force in the Landau–Lifshitz form [53] is given by

$$g^\mu = -\frac{2e^3}{3m_e c^3} \left\{ \frac{\partial F^{\mu\nu}}{\partial x^\lambda} u_\nu u_\lambda + \frac{e}{m_e c^2} \left[ F^{\mu\lambda} F_{\nu\lambda} u^\nu - F_{\nu\lambda} u^\lambda \ F^{\nu\kappa} u_\kappa \ u^\mu \right] \right\}. \tag{3}$$

Here $p^\mu$, $u^\mu$, and $x^\mu$ are the electron momentum, velocity and coordinate; $e$ and $m_e$ are the elementary charge and the electron mass; $c$ is the speed of light in vacuum. The 4-tensor of the electromagnetic field $F_{\mu\nu}$ is defined as

$$F_{\mu\nu} = \partial_\mu A_\nu - \partial_\nu A_\mu, \tag{4}$$

where $A^\mu = \varphi, \boldsymbol{A}$ is the four-vector potential with scalar $\varphi$ and vector $\boldsymbol{A}$ potentials.

Retaining the high-order terms in the limit of $\gamma_e \gg 1$, where $\gamma_e$ is the electron Lorentz factor, the three-dimensional form of the radiation friction force can be presented in the form [53]

$$\boldsymbol{g}_{LL} = -\frac{2e^4}{3m_e^2 c^4} \gamma_e^2 \frac{\boldsymbol{v}}{c} (F_{\mu\nu} u^\nu)(F^{\mu\nu} u_\nu). \tag{5}$$

where $\boldsymbol{v}$ is the electron velocity. This expression can be rewritten via the relativistic and gauge invariant parameter $\chi_e$ given by

$$\chi_e = \frac{\sqrt{-(F^{\mu\nu} p_\nu)^2}}{E_S m_e c} \tag{6}$$

where

$$E_S = \frac{m_e^2 c^3}{e\hbar}. \tag{7}$$

is equal to critical electric field of quantum electrodynamics. This field, equal to $1.32 \times 10^{18}\, \text{V/cm}$ is also known as the Schwinger field. It produces over the distance equal to the Compton wavelength, $\lambda_C = \hbar / m_e c \approx 3.86 \times 10^{-11}\, \text{cm}$, work equal to $m_e c^2$.

The parameter $\chi_e$ can be expressed via the electric and magnetic fields and electron momentum as

$$\chi_e = \frac{1}{E_S}\sqrt{\left(\gamma_e \boldsymbol{E} + \frac{1}{m_e c}\boldsymbol{p}\times\boldsymbol{B}\right)^2 - \frac{1}{m_e^2 c^2}(\boldsymbol{p}\cdot\boldsymbol{E})^2}\ . \qquad (8)$$

For an electron counter-propagating with respect to the electromagnetic wave, this parameter is approximately equal to $\chi_e \approx 2\gamma_e a_0 / a_S$, where $a_S = eE_S / m_e \omega c = m_e c^2 / \hbar\omega$ is the normalized Schwinger field. For 1-µm wavelength laser radiation, the normalized Schwinger field equals $a_S \approx 4.1\times 10^5$.

Using the expressions written above we can cast the radiation friction force as

$$\boldsymbol{g}_{LL} = -\frac{2}{3}\left(\frac{e}{\lambda_C}\right)^2 \boldsymbol{\beta}\chi_e^2 = -\frac{2\alpha m_e^2 c^3}{3\hbar}\boldsymbol{\beta}\chi_e^2, \qquad (9)$$

where $\alpha = e^2/\hbar c \approx 1/137$ is the fine structure constant and $\boldsymbol{\beta} = \boldsymbol{v}/c$ is the normalized electron velocity. In the ultra-relativistic limit, when the momentum of the electrons colliding head-on with the electromagnetic wave is well above $m_e c$, the square of the invariant parameter is approximately equal to

$$\chi_e^2 \approx \gamma_e^2 \frac{(\boldsymbol{E}+\boldsymbol{\beta}\times\boldsymbol{B})^2}{E_S^2}\ . \qquad (10)$$

The quantum effects that lead to a reduction in the rate of energy being lost to radiation can be taken into account by modifying expression (5) as

$$\boldsymbol{g} = \boldsymbol{g}_{LL} G(\chi_e), \qquad (11)$$

where the Gaunt factor $G(\chi_e)$ is equal to the ratio of the full radiation intensity to the intensity emitted by a classical electron.

Using the results published in [54], we can write the Gaunt factor $G(\chi_e)$ as

$$G(\chi_e) = -\frac{3}{4}\int_0^\infty \left[\frac{4+5\chi_e x^{3/2}+4\chi_e^2 x^3}{(1+\chi_e x^{3/2})^4}\right]\mathrm{Ai}'(z)xdx\ , \qquad (12)$$

where $z=(4x/\chi)^{2/3}$, Ai(x) is the Airy function [55]. In the following sections, we neglect the effects of the discrete nature of photon emission in quantum electrodynamics [56-58], which results in stochastic behavior of the radiating electron (see [59] and review articles [10,19] and literature cited therein). These effects will be addressed at the end of the paper.

In the limit $\chi_e \ll 1$, the form-factor $G(\chi_e)$ tends to unity as

$$G(\chi_e) = 1 - \frac{55\sqrt{3}}{16}\chi_e + \ldots = 1 - 5.9\chi_e + \ldots. \qquad (13)$$

For $\chi_e \gg 1$, it tends to zero as

$$G(\chi_e) = \frac{32\pi}{27 \times 3^{5/6}\Gamma(1/3)\chi_e^{4/3}} + \ldots = \frac{0.56}{\chi_e^{4/3}} + \ldots. \qquad (14)$$

In what follows, we shall use the approximation [60]

$$G(\chi_e) = \frac{1}{(1 + 18\chi_e + 69\chi_e^2 + 73\chi_e^3 + 5.806\chi_e^2)^{1/3}}. \qquad (15)$$

Within the interval $0 < \chi_e < 20$, the accuracy of approximation is better than 1% as it follows from the comparison of expressions (12) and (15).

Expressions (9) and (11) for the radiation friction force can be rewritten as

$$\boldsymbol{g} = -\frac{2}{3}\left(\frac{e}{\lambda_C}\right)^2 \frac{\chi_e^2}{(1 + 18\chi_e + 69\chi_e^2 + 73\chi_e^3 + 5.806\chi_e^2)^{1/3}}\boldsymbol{\beta}. \qquad (16)$$

We note that the leading term in the Landau-Lifshitz equation with a quantum correction, Eq. (11), appears from the Fokker-Planck equation obtained by [48].

## 3. Ultrarelativistic electron beam slowing down

Here we consider the head-on collision of an ultrarelativistic electron with a laser pulse. The laser ponderomotive pressure pushes the electron perpendicular to the pulse propagation direction and changes the longitudinal component of the electron momentum. The reduction in momentum due to radiation friction can be either weaker or stronger than the ponderomotive force action depending on the laser pulse amplitude, inhomogeneity, and the electron energy. In the case where the radiation friction force is negligibly weak, assuming that the electromagnetic configuration can be described by a 1D-plane wave laser pulse propagating along the *x*-axis with constant velocity, the electron dynamics are determined by the conservation of the integrals of motion [53]. They are the generalized transverse momentum

$$\boldsymbol{p}_\perp - \frac{e}{c}\boldsymbol{A}_\perp(x - ct) = \text{constant} \qquad (17)$$

and

$$m_e c^2 \gamma_e - c p_\| = (m_e^2 c^4 + p_\perp^2 c^2 + p_\|^2 c^2)^{1/2} - p_\| c = \text{constant} . \quad (18)$$

If the electron before interaction with the laser pulse has longitudinal and transverse momentum components equal to $p_{0\|} = -|p_{0\|}|$ and 0, respectively, corresponding to a head-on collision, Eqs. (17) and (18) show that

$$p_\| = -|p_{0\|}| + m_e c \frac{a^2 m_e c}{2[(m_e^2 c^2 + p_{0\|}^2)^{1/2} + |p_{0\|}|]} . \quad (19)$$

Here $a = e|A_\perp|/m_e c^2$ is the normalized vector-potential of the electromagnetic wave. As follows from Eq. (19), the longitudinal component of the electron momentum decreases. If $|p_{0\|}|$ is small,

$$|p_{0\|}| < m_e c \frac{a^2}{2(1+a^2)^{1/2}} , \quad (20)$$

the electron stops and is reflected back by the ponderomotive force. We note that the electron energy,

$$m_e c^2 \gamma_e = m_e c^2 \gamma_{e0} + m_e c^2 \frac{a^2 m_e c}{2[(m_e^2 c^2 + p_{0\|}^2)^{1/2} + |p_{0\|}|]} , \quad (21)$$

does not vanish.

The transverse scattering of the electron is taken into account in the electron rest frame, where the laser pulse duration is $\bar{\tau}_{las} \approx \tau_{las} \sqrt{1+a_0^2}/2\gamma_e$. The electron is not significantly scattered aside by the laser ponderomotive force provided that its energy is large enough,

$$\gamma_{e0} > c \tau_{las} a_0 / 2 w_\perp , \quad (22)$$

where $w_\perp$ is the laser width at focus and $a_0$ is the laser pulse amplitude. For a 1 μm wavelength, 10 PW pulse focused into a one-lambda focal spot, the normalized laser amplitude is $a_0 = 10^3$. For a 30-fs duration laser pulse this condition requires $\gamma_e > 5000$, i.e., an electron energy above 2.5 GeV. According to the condition (22) the electron is not reflected back provided that its momentum is higher than approximately $|p_{0\|}| \approx m_e c a_0$, i.e. $\gamma_e > 1000$, Further, we assume that these conditions are respected.

The electron energy changes due to the radiation losses. In the limit $\chi_e \ll 1$, where the form-factor $G(\chi_e)$ tends to unity, one can obtain from Eq. (1) an equation for the *x*-component of the electron momentum

$$\frac{dp_x}{dt} = -4\varepsilon_{rad}\omega_0 a^2(2t)\frac{p_x^2}{m_e c}, \qquad (23)$$

where it is assumed the head-on relativistic electron collision with the laser pulse depending on time and coordinate as $a(t - x/c)$. Here and below we use the dimensionless parameter

$$\varepsilon_{rad} = \frac{2e^2\omega}{3m_e c^3} = \frac{4\pi r_e}{3\lambda} \qquad (24)$$

with classical electron radius $r_e = e^2/m_e c^2 \approx 2.82\times 10^{-13}$ cm and the laser wavelengh $\lambda$. For one-micron laser wavelength the the parameter $\varepsilon_{rad}$ approximately equals $1.18\times 10^{-8}$. The solution of Eq. (23) is given by

$$p_x(t) = \frac{p_x(0) m_e c}{m_e c + 4\varepsilon_{rad}\omega_0 p_x(0)\int_0^t a^2(2t')dt'}. \qquad (25)$$

From this expression it is seen that the radiation time of the electron energy loss is

$$\tau_{rad,1} = \frac{m_e c}{4\varepsilon_{rad}\omega_0 p_x(0) a_0^2}. \qquad (26)$$

It can be written in the form $\tau_{rad,1} = T_0/8\pi\varepsilon_{rad}\omega_0\gamma_0 a_0^2$, where $T_0 = 2\pi/\omega_0$ is the electromagnetic wave period, $\gamma_0 \approx p_x(0)/m_e c$ is the electron gamma-factor, and $a_0$ is the laser pulse amplitude. For one-micron wavelength laser pulse the radiation time is approximately equal to $\tau_1 \approx 3T_0 10^6/a_0^2\gamma_{b,0}$ with $T_0 \approx 3$ fs. The radiation loss effects are relatively weak if $a_0^2\gamma_0 < 3\times 10^6$, e. g. if the 500 MeV electron interacts with the laser pulse with the intensit lower than $I = 4\times 10^{21}$ W/cm$^2$.

If we assume an $a(t)$ dependence of the form $a(t) = a_0 \exp(-t^2/2\Delta t^2)$, the expression for $p_x(t)$ can be rewritten as

$$p_x(t) = \frac{p_x(0) m_e c}{m_e c + \sqrt{\pi}\varepsilon_{rad} p_x(0)\omega_0\Delta t a_0^2 \text{erf}(2t/\Delta t)}. \qquad (27)$$

Here $\text{erf}(x)$ is the error function equal to [54]

$$\text{erf}(x) = \frac{2}{\sqrt{\pi}} \int_0^x \exp(-t^2) dt. \tag{28}$$

Equation (25) shows that for large enough $p_x(0)\tau a_0^2$, the electron momentum tends to the limit of

$$p_x(t) \underset{t \to \infty}{\to} \frac{m_e c}{\sqrt{\pi} \varepsilon_{rad} \omega_0 \Delta t a_0^2} \tag{29}$$

in accordance with the theory formulated in the book by L. D. Landau and E. M. Lifshitz [53]. For $a_0 = 10^2$ and $\omega_0 \Delta t = 6$, where the frequency $\omega_0$ corresponds to the wavelength $\lambda_0 = 2\pi c/\omega_0 = 1\mu m$, the normalized electron momentum $p_x(\infty)$ is approximately equal to $500 m_e c$, i.e. for a single-cycle one-micron wavelength laser with an intensity of $\approx 10^{22}$ W/cm$^2$ the electron energy is approximately equal to 250 MeV.

In the limit where quantum corrections weaken radiation friction, i.e., when according to Eq. (15) the Gaunt factor is approximately equal to $G(\chi_e) \approx 0.56/\chi_e^{4/3}$, the equation of the electron motion with the radiation friction force given by Eq. (1) can be written in the form

$$\frac{dp_x}{dt} = -\eta \, \varepsilon_{rad} \omega_0 m_e c a_0^{2/3}(2t) a_S^{4/3} \left(\frac{p_x}{m_e c}\right)^{2/3} \tag{30}$$

with the dimensionless coefficient approximately equal to $\eta \approx 0.888$. Here, the dimensionless parameter $a_S = eE_S/m_e \omega_0 c = m_e c^2/\hbar \omega_0$ is the normalized Schwinger field. The solution to the equation (30) is

$$p_x = p_{x,0} \left[1 - \frac{\eta \varepsilon_{rad} \omega_0 m_e^{1/3} c^{1/3} a_S^{4/3}}{3 p_{x,0}^{1/3}} \int_0^t a_0^{2/3}(2t) dt \right]^3. \tag{31}$$

For a constant amplitude laser pulse the electron momentum formally tends to zero during the radiation time

$$\tau_{rad,2} = \frac{3 p_{x,0}^{1/3}}{\eta \, \varepsilon_{rad} \omega_0 m_e^{1/3} c^{1/3} a_S^{4/3} a_0^{2/3}}, \tag{32}$$

which can be written as $\tau_{rad,2} = 4\tau_{rad,1} \chi_e^{-4/3}/\eta$.

## 4. Change of the electron energy spectrum.

Radiation losses lead to a change in the energy spectrum of the electrons interacting with a strong localized electromagnetic field. The evolution of the electron energy distribution can be described by the kinetic equation

$$\partial_t f + \partial_{p_x}(A(t, p_x) f) = 0, \qquad (33)$$

where

$$A(t, p_x) = \dot{p}_x, \qquad (34)$$

with the radiation friction force, $\dot{p}_x$, given by either Eq. (23) in the limit $\chi_e \ll 1$ or by Eq. (30) when $\chi_e \gg 1$. The solution to Eq. (33) is the function $f(t, p_x)$. This function is constant on the characteristics (e.g., see [61]). The equations for the characteristics of Eq. (33) are

$$dt = \frac{dp}{\dot{p}_x} = \frac{-df}{f \, \partial_{p_x} \dot{p}_x}. \qquad (35)$$

In the cases of the radiation friction force given by Eqs. (23) and (29), the radiation friction force can be represented in the form

$$\dot{p}_x = \upsilon(t) \varpi(p_x). \qquad (36)$$

Here, the functions $\upsilon(t)$ and $\varpi(p_x)$ are

$$\upsilon(t) = -4 \varepsilon_{rad} \omega_0 m_e c a^2(2t) \quad \text{and} \quad \varpi(p_x) = \left(\frac{p_x}{m_e c}\right)^2 \qquad (37)$$

in the case corresponding to Eq. (23) and

$$\upsilon(t) = -\eta \, \varepsilon_{rad} \omega_0 m_e c a^{2/3}(2t) a_S^{4/3} \quad \text{and} \quad \varpi(p_x) = \left(\frac{p_x}{m_e c}\right)^{2/3} \qquad (38)$$

for the radiation friction force given by Eq. (30), respectively. Introducing the function

$$F = \varpi f \qquad (39)$$

and changing the variables to

$$s = -\int^t \upsilon(t') dt' \quad \text{and} \quad w = \int^{p_x} \frac{dp}{\varpi(p)}, \qquad (40)$$

we rewrite the kinetic equation (33) as

$$\partial_s F - \partial_w F = 0. \qquad (41)$$

The initial value problem solution of this equation is

$$F(s, w) = F_0(w - s) \qquad (42)$$

with $F_0(w)$ determined by the initial conditions at $s=0$. For the distribution function $f(t,p_x)$ this yields the expression

$$f(t,p_x) = \frac{\varpi(p_{x,0})}{\varpi(p_x)} f_0(p_{x,0}), \qquad (43)$$

where $f_0(p_{x,0})$ is the distribution function at $t=0$.

As an example, we consider the electron distribution function prior to the electron beam interacting with an electromagnetic field of the super-Gaussian form

$$f_0(p_{x,0}) = \frac{1}{2\Gamma(1+1/m)\Delta p_0} \exp\left[-\left(\frac{p_{x,0}-p_{b,0}}{\Delta p_0}\right)^m\right] \qquad (44)$$

with positive index $m$. Here $\Gamma(x)$ is the Gamma function [55]. This distribution function describes an electron beam with average momentum $p_{b,0}$ and a width in the momentum space equal to $\Delta p_0$. Fig. 1 shows electron beam distribution functions for $m=2,4,6,8$ and $p_{b,0}=3$, and $\Delta p_0 = 1$. Here, we use the momentum normalization to

$$\Pi_1 = \frac{m_e c}{4\varepsilon_{rad} a_0^2}. \qquad (45)$$

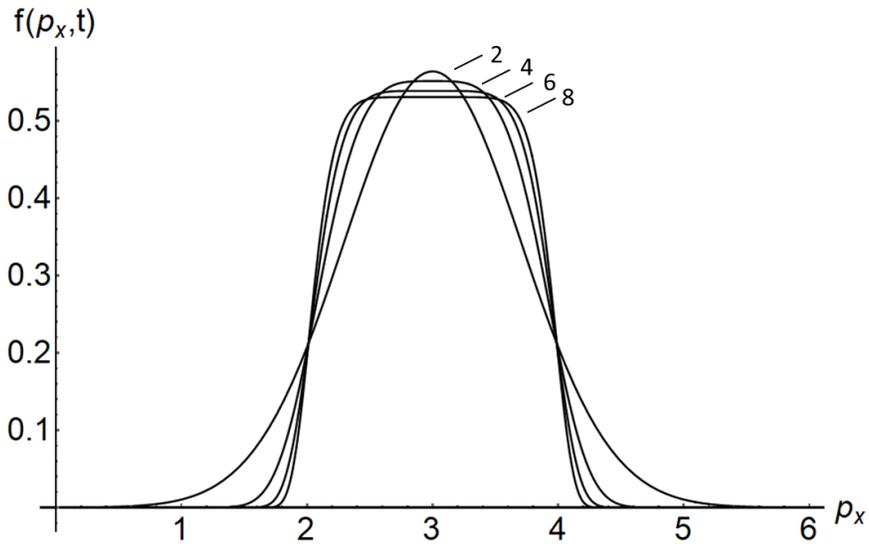

Fig. 1. Electron beam distribution functions (44) for $m=2,4,6,8$ and $p_{b,0}=3$, and $\Delta p_0 = 1$. Here we use the momentum normalization to $\Pi_1$ given by Eq. (45).

Using the relationships presented above, we obtain in the case where the radiation friction force $\dot{p}_x$ is given by Eqs. (36,37), that the electron distribution function depends on time as

$$f_1(t, p_x) = \frac{1}{2\Gamma(1+1/m)\Delta p_0 \Pi_1 (1-p_x s_1 t)^2} \exp\left\{-\left[\frac{p_x - p_{b,0}(1-p_x s_1 t)}{\Delta p_0 (1-p_x s_1 t)}\right]^m\right\}, \quad (46)$$

where

$$s_1 = \frac{\omega_0}{\Pi_1} \quad (47)$$

with $\Pi_1$ given by Eq. (45). According to this expression, the average electron momentum and the distribution width decrease as

$$p_b(t) = \frac{p_{b,0}}{1 + p_{b,0} s_1 t} \quad \text{and} \quad \Delta p(t) \approx \frac{\Delta p_0}{1 + p_{b,0} s_1 t}. \quad (48)$$

Fig. 2 shows the electron beam distribution functions given by Eq. (46) for $m = 8$ at $s_1 t = 0.0, 0.1, 0.2, 0.3, 0.4, 0.5$. For the sake of simplicity, here we assume that the normalized wave amplitude is constant and equal to $a_0$, thereby corresponding to circularly polarized radiation. As is clearly seen, the radiation friction effects result in a reduction of the average electron momentum and narrowing of the distribution.

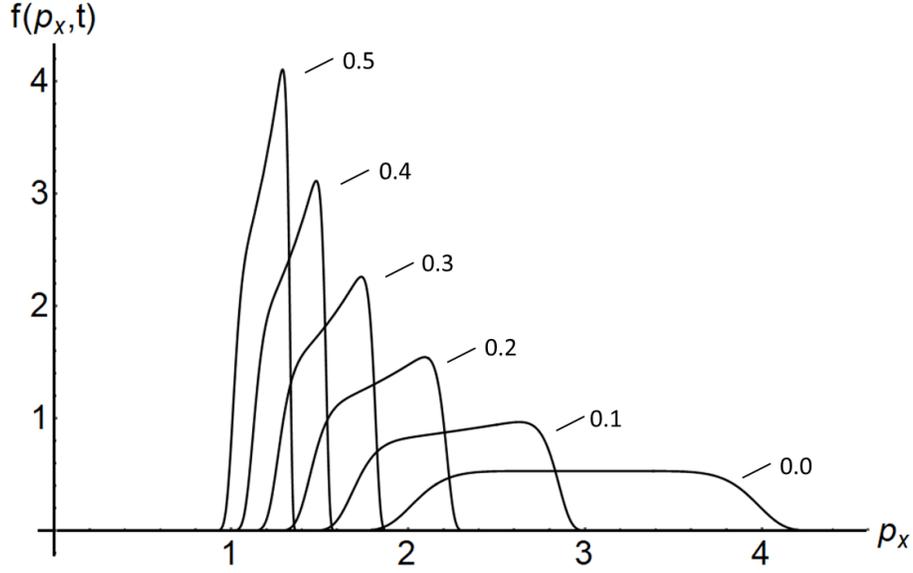

Fig. 2. Electron beam distribution functions (46) versus time for $m=8$, $p_{b,0}=3$ and $\Delta p_0=1$ at $s_1 t = 0.0, 0.1, 0.2, 0.3, 0.4, 0.5$. The electron momentum is normalized to $\Pi_1$.

We note that in this case the characteristic time of the energy loss is approximately equal to the time given by Eq. (26). Fig. 2 corresponds to relatively short energy loss time, i. e. to strong radiation friction effects.

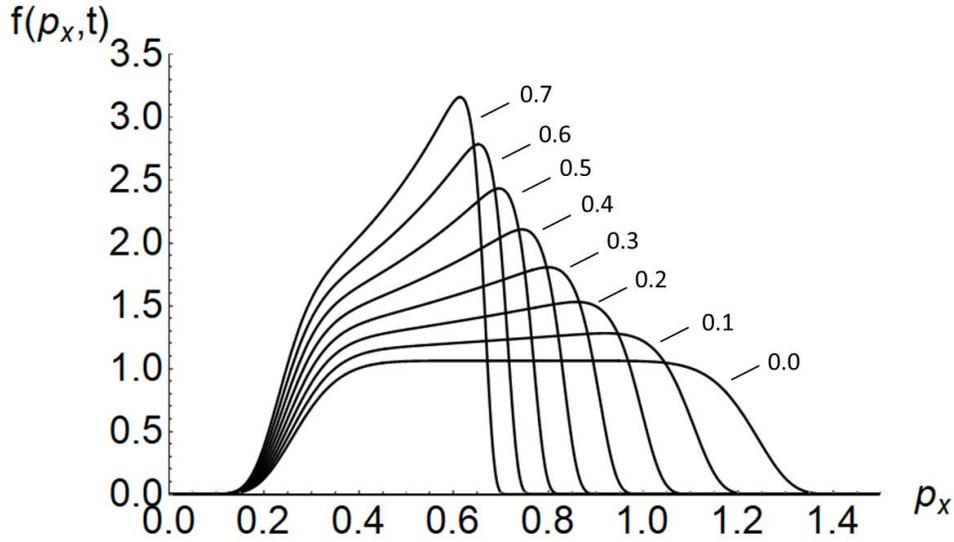

Fig. 3. Electron beam distribution functions (46) versus time for $m=8$, $p_{b,0}=0.75$ and $\Delta p_0 = 0.5$ at $s_1 t = 0.0, 0.1, 0.2, 0.3, 0.4, 0.5, 0.6, 0.7$. The electron momentum is normalized to $\Pi_1$.

In Fig. 3 we show the electron beam distribution functions given by Eq. (46) for $m = 8$, $p_{b,0} = 0.75$ and $\Delta p_0 = 0.5$ at $s_1 t = 0.0, 0.1, 0.2, 0.3, 0.4, 0.5, 0.6, 0.7$, which corresponds to relatively long energy loss time, i. e. to weak radiation friction effects.

In the case where the radiation friction force is given by Eqs. (36, 38), the electron distribution function is

$$f_2(t, p_x) = \frac{1}{2\Gamma(1+1/m)\Delta p_0 \Pi_2} \left(1 + \frac{s_2 t}{p_x^{1/3}}\right)^2 \exp\left\{-\left[\frac{\left(p_x^{1/3} + s_2 t\right)^3 - p_{b,0}}{\Delta p_0}\right]^m\right\} \quad (49)$$

where the electron momentum is normalized by

$$\Pi_2 = \frac{3}{\eta \varepsilon_{rad} m_e^{1/3} c^{1/3} a_0^{2/3} a_s^{4/3}}. \quad (50)$$

and

$$s_2 = \frac{\omega_0}{\Pi_2} \quad (51)$$

For the sake of simplicity here we assume that the normalized wave amplitude is constant and equal to $a_0$, again corresponding to circularly polarized radiation. Fig. 4 shows the electron beam distribution functions, $f_2(t, p_x)\Pi_2$, given by Eq. (49) for $m = 8$, at $s_2 t = 0.0, 0.1, 0.2, 0.3, 0.4, 0.5$. Here, we have again assumed that the normalized wave amplitude is constant and equal to $a_0$. In this case, radiation friction effects also result in a reduction of the average electron momentum and narrowing of the distribution.

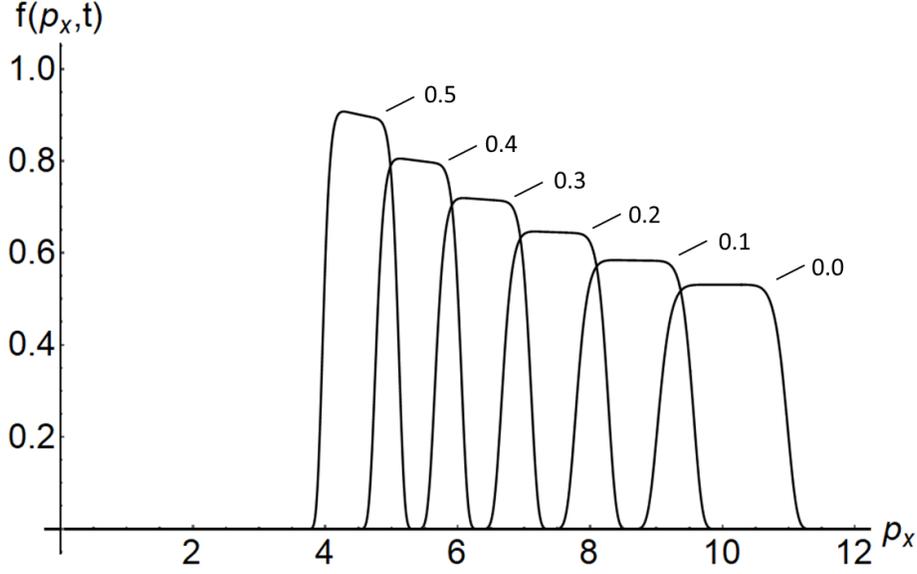

Fig. 4. Electron beam distribution function given by Eq. (49) versus time for $m = 8$, $p_{b,0} = 10$ and $\Delta p_0 = 1$ at $s_2 t = 0.0, 0.1, 0.2, 0.3, 0.4, 0.5$. The electron momentum is normalized on $\Pi_2$.

In the case of distribution function $f_1(t, p_x)$ shown in Fig. 2, the spectral narrowing is faster than in the case of the function $f_2(t, p_x)$ presented in Fig. 4. The characteristic time of the energy loss is approximately equal to the time given by Eq. (32).

To further exemplify these effects, we consider an initial electron distribution function having a form typical for laser wake field accelerated electrons. As shown in [62, 63] the shape of the energy spectrum of electrons accelerated by wake-fields [64] can be approximated by the formula

$$\left.\frac{dN}{d\mathcal{E}}\right|_{\mathcal{E} \to \mathcal{E}_m - 0} = \frac{2N_0}{\pi \sqrt{\mathcal{E}_m^2 - \mathcal{E}^2}}, \qquad (52)$$

where $\mathcal{E}_m$ is the electron maximal energy assuming that $\mathcal{E}_m > \mathcal{E}$. Correspondingly, the electron distribution function prior to interaction with the electromagnetic field is

$$f_0(p_{x,0}) = \frac{2}{\pi \sqrt{p_{m,0}^2 - p_{x,0}^2}}. \qquad (53)$$

with $p_m$ being the maximal electron momentum. This distribution has an integrable singularity at

$p = p_m$.

In the case where $\dot{p}_x$ is given by Eqs. (36, 37) the electron distribution function depends on time as

$$f(p_x,t) = \frac{2}{\pi \Pi_1 (1 - p_x s_1 t)\sqrt{p_{m,0}^2 (1 - p_x s_1 t)^2 - p_x^2}}. \tag{54}$$

It is plotted in Fig. 5 for $s_2 t = 0.0, 0.1, 0.2, 0.3, 0.4, 0.5$ and $p_{m,0} = 2.5$. In this case, the radiation friction effects also result in decreasing maximum electron momentum as

$$p_m(t) = p_{m,0}/(1 + p_{m,0} s_1 t). \tag{55}$$

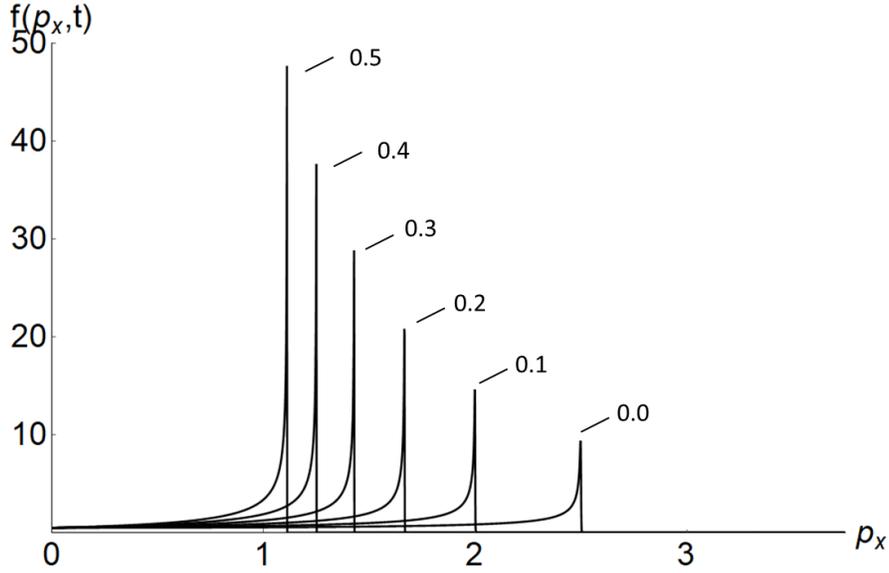

Fig. 5. Electron beam distribution function given by Eq. (54) versus time for $p_{m,0} = 2.5$ at $s_1 t = 0.0, 0.1, 0.2, 0.3, 0.4, 0.5$. The electron momentum is normalized on $\Pi_1$.

In Fig. 6 we show the electron distribution function when radiation friction is described by equations (36, 38). It is given by

$$f(p_x,t) = \frac{2(1 + s_2 t / p_x^{1/3})^2}{\pi \Pi_2 \sqrt{p_{m,0}^2 - (p_x^{1/3} + s_2 t)^6}}. \tag{56}$$

The electron beam distribution is presented for $p_{m,0} = 15$ at $s_2 t = 0.0, 0.1, 0.2, 0.3, 0.4, 0.5, 0.6$. According to expression (54), the maximum electron momentum decreases as

$$p_m(t) = (p_{m,0}^{1/3} - s_2 t)^3. \tag{57}$$

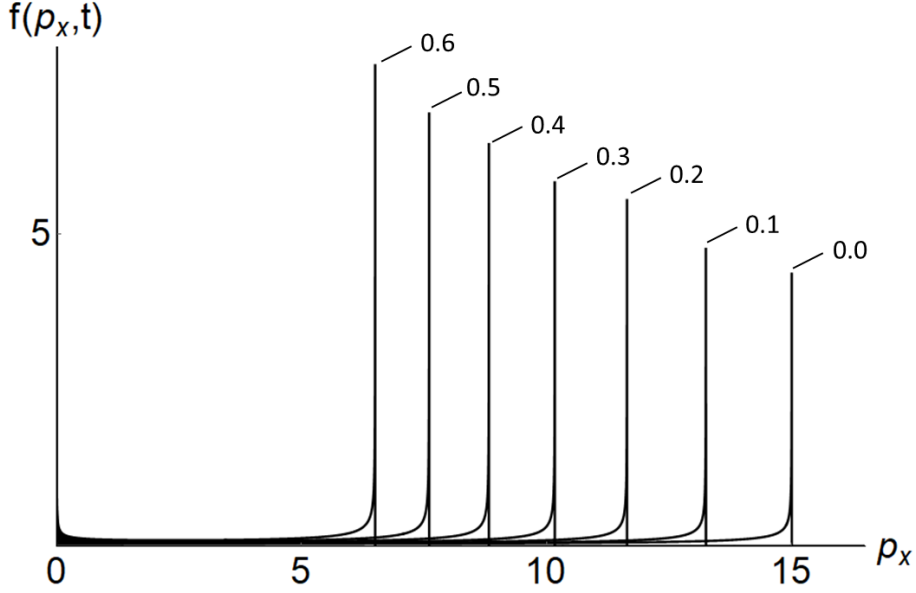

Fig. 6. Electron beam distribution function given by Eq. (56) versus time for $p_{m,0} = 15$ at $s_2 t = 0.0, 0.1, 0.2, 0.3, 0.4, 0.5, 0.6$. The electron momentum is normalized on $\Pi_2$.

As we see in the case of the electron beam with an initial energy spectrum described by the super-Gaussian function (44), radiation cooling results in an overall reduction in the mean energy of the electron distribution, however, asymmetry arises in the distribution because the energy loss is greater for the higher energy part of the spectrum. This is clearly seen in Figs. 2 and 4. When the pre-collision electron energy spectrum has the form typical for a LWFA electron beam (it is inversely proportional to the square root of $\sqrt{p_m - p}$ as given by Eq. (53)), radiation cooling does not change the type of singularity, leading to a reduction of the maximum electron momentum according to expressions (55-57).

## 5. Electron energy spectrum broadening due to quantum mechanical stochasticity effects

As we discussed above, within the framework of the approximation based on classical electrodynamics, the expressions for the radiation friction force, including the case when the quantum effects are taken into account with the Gaunt factor $G(\chi_e)$, radiation losses for electrons interacting with the electromagnetic wave result in a drift of the electron distribution function towards lower electron momentum with additional narrowing of the momentum distribution function. The quantum mechanical stochasticity leads to the appearance of diffusion in the momentum space. The Fokker-Planck equation implementing energy drift and diffusion takes the form [65]

$$\partial_t f = \partial_{p_x}\left\{A(p_x)f + \frac{1}{2}\partial_{p_x}\left[B(p_x)f\right]\right\}. \tag{58}$$

It is convenient to write the energy drift and diffusion coefficients as

$$A(p_x) = \frac{2\alpha m_e^2 c^3}{3\hbar}\beta\chi_e^2 \quad \text{and} \quad B(p_x) = \frac{55}{8\sqrt{3}}\frac{\alpha m_e^3 c^4}{\hbar}\gamma\chi_e^3. \tag{59}$$

Estimating characteristic energy drift and diffusion time as

$$\tau_{drift} = p_x/A(p_x) \quad \text{and} \quad \tau_{diff} = 2p_x^2/B(p_x) \tag{60}$$

we find that the energy drift evolves faster than the energy diffusion, $\tau_{drift} \ll \tau_{diff}$, at relatively low electron energy: $\gamma < 8/55\sqrt{3})(a_S/a_0) \approx 8\times 10^5/a_0$ (see also discussion in [59, 65]). If the normalized field amplitude equals $a_0 = 10^3$, which can be reached with multi-petawatt lasers, electron energy $8\times 10^2 m_e c^2$ corresponds to 400 MeV. Taking into account that the QED parameter equals $\chi_e = (a_0/a_S)\gamma$ we find that the energy diffusion is faster than the energy drift for $\chi_e > 1/3$.

Introducing normalized time and momentum, $t/\text{T} = \tau$ and $p_x/\Pi = p$, with

$$\text{T} = \frac{3^{3/2}55}{128}\frac{\hbar}{\alpha m_e c^2}\left(\frac{a_S}{a_0}\right) \quad \text{and} \quad \Pi = \frac{16}{55\sqrt{3}}m_e c\left(\frac{a_S}{a_0}\right), \tag{61}$$

we can rewrite Eq. (58) in the form

$$\partial_\tau f = \partial_p\left\{(p+\bar{\gamma})^2\frac{p}{\bar{\gamma}}f + \partial_p\left[\bar{\gamma}(p+\bar{\gamma})^3 f\right]\right\} \tag{62}$$

with

$$\bar{\gamma} = \sqrt{\left(\frac{m_e c}{\Pi}\right)^2 + p^2}. \tag{63}$$

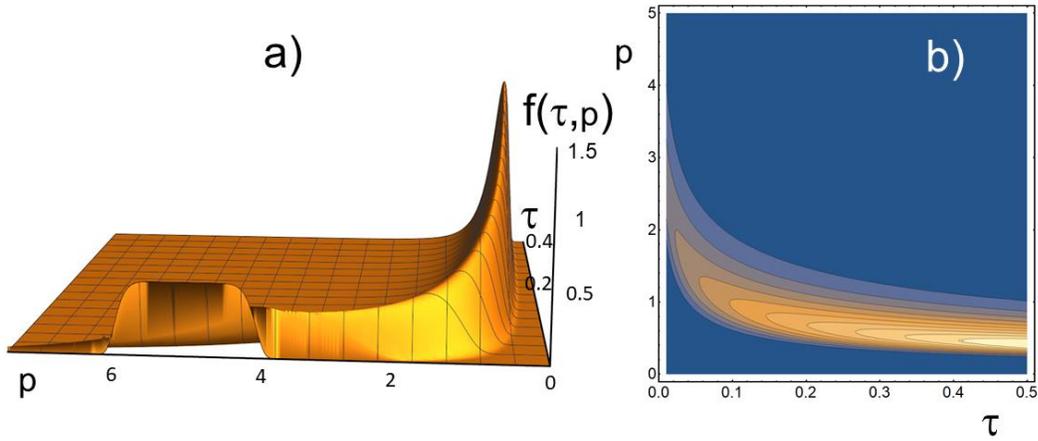

Fig. 7. Electron beam evolution for initial distribution function given by Eq. (44) with $m = 8, p_0 = 5, \Delta p_0 = 1$; a) function $f(\tau, p)$; b) equal value contours of $f(\tau, p)$ on the plane $(\tau, p)$. The electron momentum is normalized on $\Pi$ given by Eq. (61).

In Fig. 7 we show the results of the electron beam evolution with an initial distribution function given by Eq. (44) with $m = 8, p_0 = 5, \Delta p_0 = 1$. In frame a) we present the function $f(\tau, p)$. Frame b) shows equal value contours of $f(\tau, p)$ on the plane $(\tau, p)$. From this, we observe an overall drift of the energy spectrum towards lower energies. Diffusion then leads to asymmetric spectral broadening, which is stronger in the high-energy wing of the distribution. At the very late stage, radiation friction results in energy distribution narrowing.

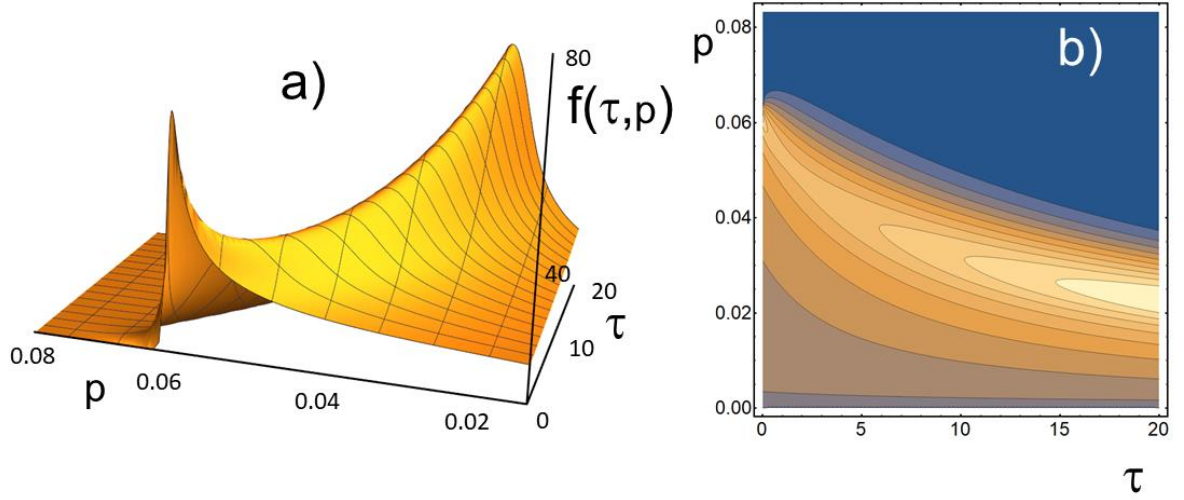

Fig. 8. Electron beam evolution for initial distribution function given by Eq. (53) with $p_0 = 0.0625$; a) function $f(\tau, p)$; b) equal value contours of $f(\tau, p)$ on the plane $(\tau, p)$. The electron momentum is normalized on $\Pi$.

Fig. 8 presents the results of the electron beam evolution for the initial distribution function given by Eq. (53) with $p_0 = 0.0625$. Frame a) shows the function $f(\tau, p)$. In Frame b) we plot constant value contours of $f(\tau, p)$ on the plane $(\tau, p)$. We again observe a systematic drift of the energy spectrum towards low energies due to radiation losses. Diffusion effects result in a smoothing of the distribution in the region of vicinity of the maximum electron momentum and broadening of the whole momentum distribution.

Asymptotically at late times, the electron distribution tends to a stationary state described by the solution of the equation (61) i.e. of the equation

$$\partial_p \left( B(p) f_s \right) + 2A(p) f_s = 0. \qquad (64)$$

The right-hand side of this equation is assumed equal to zero which corresponds to vanishing momentum flux. Its solution has the form

$$f_s(p) = U(p) \exp[\Psi(p)] \qquad (65)$$

with

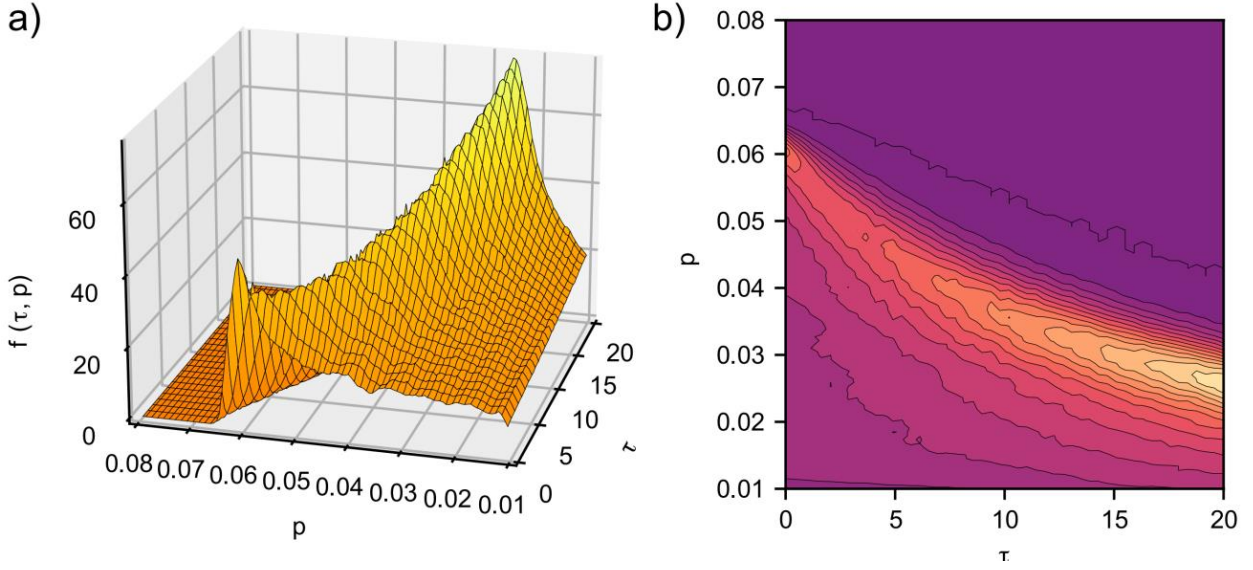

Fig. 9. Ptarmigan simulated evolution of the electron distribution used in Fig. 8, constructed using $10^4$ particles. The electron beam was collided head-on with a circularly polarized $a_0 = 10$ plane wave.

$$U(p_x) = \frac{\varepsilon_1^4 \left(p\sqrt{\varepsilon_1^2 + p^2} + \varepsilon_1^2 + p^2\right)^3}{\left(\varepsilon_1^2 + p^2\right)^2 \left(\sqrt{\varepsilon_1^2 + p^2} + p\right)^6}, \quad (66)$$

where $\varepsilon_1 = m_e c / \Pi$, and

$$\Psi(p) = \frac{p - \sqrt{\varepsilon_1^2 + p^2}}{\varepsilon_1^2} + \frac{\text{ArcTan}(p/\varepsilon_1)}{\varepsilon_1}. \quad (67)$$

In the limit $p \to 0$ for these functions we have

$$U(p_x) = \frac{1}{\varepsilon_1^7} - \frac{5p}{\varepsilon_1^8} + \frac{11p^2}{\varepsilon_1^9} + O[p]^3, \quad (68)$$

$$\Psi(p) = -\frac{1}{\varepsilon_1} + \frac{2p}{\varepsilon_1^2} - \frac{p^2}{2\varepsilon_1^3} + O[p]^3. \quad (69)$$

When $p \to \infty$ the functions have the order

$$U(p_x) = \frac{\varepsilon_1}{32 p^8} + O\left[\frac{1}{p}\right]^{10}, \quad (70)$$

$$\Psi(p) = \frac{\pi}{2\varepsilon_1} + \frac{3}{2p} + O\left[\frac{1}{p}\right]^3. \tag{71}$$

## 6. Simulation results from the Monte Carlo code

In order to benchmark the Fokker-Planck equation approach used in the previous section to account for quantum mechanical stochasticity effects, we employ a Monte Carlo code Ptarmigan [67, 68] to model electron beam collision with a laser pulse. The Ptarmigan code is a single particle code, which takes into account SFQED effects (multi photon Compton and Breit-Wheeler processes) when charged particles propagate in strong electromagnetic fields. These effects are described using either a local constant field approximation (LCFA) or local monochromatic approximation (LMA). The photon emission and pair production are treated as point-like events, which modify particle 4-momentum and create new particles. Between these events, the particle motion is treated according to classical equations of motion in in electromagnetic field. The Ptarmigan code takes into account the angular distribution of secondary particles in multi-photon Compton and Breit-Wheeler processes, which is different from the distribution that can be observed in typical PIC-QED codes, where the collinear emission approximation is used (see [10] for details). The results of the Ptarmigan modeling of an electron beam with the same momentum distribution as used in Fig. 8 (using $10^4$ particles) with a circularly polarized electromagnetic wave with $a_0 = 10$ are shown in Fig. 9.

Here, we employed the local constant field approximation (LCFA), since the energy of electrons and electromagnetic field strength allow this. Fig. 9 shows the evolution of the electron distribution, which has been smoothed to reduce noise. The electron beam evolution is very similar to the numerical results obtained by solving the diffusion equations, which are shown in Fig. 8. The distribution initially broadens, followed by narrowing at late times. This indicates the validity of using the diffusion equation approach to analyze the evolution of electron beam distribution during the interaction with an intense laser pulse.

## 7. Conclusion

Analysis of radiation friction effects shows an overall drift of the electron momentum distribution function towards lower momentum. This down-drift results in a narrowing of the momentum distribution accompanied by formation of an asymmetric momentum distribution. The stochastic nature of electron radiation in the quantum limit is demonstrated by a broadening of the electron distribution. The diffusion effects result in a smoothing of the distribution in the region near the maximum electron momentum, and broadening of the whole momentum distribution when the radiation losses are balanced by diffusion in the momentum space, as is the case of standard accelerators of charged particles [2,3].

**Credit Author Statement**

S. V. Bulanov: idea and theoretical model propose and results discussion

G. M. Grittani: work on the parameters relevant to coming experiments and results discussion

R. Shaisultanov: theoretical model developing and results discussion

T. Zh. Esirkepov: theoretical model developing and results discussion

C. P. Ridgers: computational and theoretical model developing and results discussion

S. S. Bulanov: theoretical model developing and results discussion

B. K. Russell: computer simulation and results discussion

A. G. R. Thomas: theoretical model developing and results discussion

All the authors participate writing - reviewing and editing

**Declaration of Competing Interest**

The authors declare the following financial interests/personal relationships, which may be considered as potential competing interests: No

**Data availability**

Data will be made available on request.

**Acknowledgement**

This work was supported by the National Science Foundation and Czech Science Foundation under NSF-GACR collaborative grant 2206059 (and 22-42963L) and NSF grant 2108075 and by the project 'Advanced Research Using High Intensity Laser Produced Photons and Particles' (ADONIS) CZ.02.1.01/0.0/0.0/16_019/0000789 from European Regional Development Fund. The authors are thankful to C. Keitel, A. Di Piazza, L. O. Silva, P. Bilbao, T. Silva, M. Vranic, T. Grismayer, T. M. Geong, M. Jirka, and P. Hadjisolomou for fruitful discussions.